\documentclass[aps,showkeys]{revtex4}
\usepackage{times}
\usepackage{epsfig}
\usepackage{amssymb}
\usepackage{amsmath}
\usepackage{amsfonts}

\begin{document}

\title{The beginning of nonlinear stage of evolution of protostars at $ z=20 $}

\author{V. K. Dubrovich}
\affiliation{Special Astrophysical Observatory, St. Petersburg Branch, Russian Academy of Sciences, 196140, St. Petersburg, Russia}
\email[]{dvk47@mail.ru}

\begin{abstract}
The results of the EDGES (Experiment to Detect the Global EoR Signature) experiment (Bowman et al., 2018) is interpreted as the beginning of compression stage of primary density fluctuations in a mini halo. Estimates of the mass of these objects are given. 
\end{abstract}
\maketitle

\section{Introduction}
The evolution of the early Universe is widely studied in all directions - from the moment of the Big Bang to the era of reionization and formation of the first stars. At each stage, one have to use the most optimal models of physical processes, describing the observed effects. In particular, for the testing of the reionization epoch, one of such processes can be the effect of the formation of CMB spectral distortions in the 21 cm line of hydrogen. Other effects are also possible. But here we analyze the latest result of the EDGES experiment \cite{bowman2018}. The analysis of the experiment technique is outside of present study and we take the result as a realistic one.

The possibility of formation of this type of absorption line in the model of a homogeneous expanding universe was considered in \cite{varsh77} for the first time. In \cite{varsh77} it was shown that the temperature of the matter decreases with respect to the CMB temperature at redshifts $150> z> 15$.  In addition, the spin temperature of the ground state of the hydrogen atom was calculated, and an absorption spectrum was obtained. It turned out that the position of the minimum corresponds to $z\approx 40$ and it weakly depends on the density of matter.

The expression for brightness temperature is given by
\begin{eqnarray}
T_{b}=\frac{3c^3\hbar A_{10}n_{H}x_{HI}}{16\nu^2_{10}k_{B}(1+z)^2(dv_{||}/dr_{||})}\left(1-\frac{T_{\gamma}}{T_{s}} \right),
\end{eqnarray}
where $ A_{10} $ is the Einstein coefficient for hyperfine transition, $ n_{H} $ is the total density of hydrogen atoms and ions, $ x_{HI} $ is the ionization fraction of hydrogen, $ \nu_{10} $ is the frequency of hyperfine transition, $ dv_{||}/dr_{||} $ is the velocity gradient of matter, $ T_{\gamma} $ is the CMB radiation temperature, $ T_{s} $ is the spin temperature of the hyperfine structure of hydrogen.

In all the works devoted to the calculations of the absorption spectrum, only processes of formation of the spin temperature are taken into account. The magnitude of the velocity gradient in this case is determined only by the Hubble expansion. However, as it turns out, the gradient of velocity and metrics becomes the dominant one in the epoch before the beginning of formation of primary stars and powerful energy release (including the emission in Ly$ _{\alpha} $ lines).

\section{Formation of absorption in a homogeneous universe}

Let us consider the process of absorption of photons in an atomic line. A single photon, with some initial frequency, moves in a straight line and gradually turns red due to the expansion of space. If the matter (hydrogen) does not have inhomogeneities, then it does not have a peculiar velocity and expands uniformly. In the case when photon has an initial frequency greater than the resonance frequency $ \nu_{0} $, it falls into the Doppler transition line and interacts with matter.

In particular, this photon can be absorbed by an atom. Then the excited state can be removed by a collision with another atom. In this case the initial photon is destroyed

It is very important that the probability of photon absorption is proportional to the time $ \Delta t $ of finding a photon in the line contour - $ \nu_{D} $. For a uniform expansion, $ \Delta t/ t_0 \approx\nu_D / \nu_0$ , where $t_0$ is the age of the Universe at the given moment. In the absence of the expansion of the Universe, only photons that are initially in the line contour can interact with matter, but the time of such interaction does not depend on the width of the contour.

\section{The role of inhomogeneity}

The situation differs in the presence of inhomogeneity of the matter distribution. In this case, the ratios of the velocity gradient which is is not of gravitational origin (for example, in the collision of two fluctuations), the growth of the gravitational potential on the scale of the inhomogeneity $L$ and the dynamics of its evolution become important.

The role of the velocity gradient can be understood from simple considerations: as the magnitude of the density fluctuations increases, its expansion slows down and the compression begins gradually under the action of self-gravitation. As a result, at some instant in the region of the order of the fluctuation radius the effects of the peculiar motion of matter (Hubble expansion and self-gravitation) are compensated. This leads to an increase of the photon path length within the line contour. As a consequence, the effective optical thickness increases on the scale of the given fluctuation. The calculation of the optical thickness in the lines of primary molecules was done in \cite{dubr} but without the accounting for the self-gravitation. 

For the first time, effects of self-gravitation for some type of fluctuations in the early Universe, was consdered by Ya. Zeldovich in \cite{zeld}. A more detailed description of various variants of the nonstationary spherically symmetric metric is given in \cite{gubanov}.

\section{Numerical estimations}

In our case, the temperature difference between matter and radiation and some absorption amplitude occurs up to $z=10$ \cite{varsh77} even without primary stars and Lyman lines. This absorption has a gently sloping spectrum and magnitude about $0.1$ K at: $z\approx 20$ and an average matter density $\Omega_{b}\approx0.1$.

The shape of the spectral line is determined by the dynamics of the velocity gradient and metric. Thus, the enhancement of absorption occurs due to increase of the mean free path of photons $ L_{\rm max} $. The depth of the line is proportional to $ L_{\rm max} $.  For estimates one can take the following values of parameters: $ z\approx 20 $, $ T_{b}=10 $ K, $ t_{0}=5\times 10^{-15} $ sec, $ \rho_{0}=2\times10^{-25} $ g cm$ ^{-3} $.

Then $ \Delta \nu_{D}/\nu_{0}\approx 10^{-6} $, $ \Delta t\approx5\times 10^9 $, $ L_{D}=c \Delta t \approx 1.5\times 10^{20} $, $ M_{D}\approx 100 M_{S} $, where $ M_{S} $ is the mass of sun. To match the observed line depth (about $ 0.5 $ K, \cite{bowman2018}) with a theoretical value $ 0.1-0.2 $ K \cite{varsh77} one need to obtain amplification coefficient
\begin{eqnarray}
K=\frac{L_{\rm max}}{L_{D}}.
\end{eqnarray}
This means that the mass of the matter fluctuations  $M$ begins to contract and gives the desired amplification effect. It should be
\begin{eqnarray}
M=K^3M_{D}\approx 10^4 M_{S}.
\end{eqnarray}
Accurate calculation of the geometry and dynamics of this process can lead to an increase in the required mass by about an order of magnitude, which, nevertheless, is completely within the framework of the standard mass spectrum on a small scale.

\section{Conclusion}

To explain the spectrum, in  \cite{bowman2018} it is proposed to take into account the role of velocity gradients of matter and metric in proto-objects at the early stage of their evolution (the beginning of the nonlinear stage of compression of protostars with a mass of about $ 10^5M_{S} $ at $ z\approx 20 $). The red absorption wing is formed at the beginning of the deceleration of the cloud expansion due to a decrease in the velocity gradient and, correspondingly, an increase of the photon path length (an increase in the optical thickness). The blue wing is formed by the initiation of active compression and an increase of the velocity gradient. 

The depth of the absorption line is determined by the characteristic scale of the proto object and the details of the compression dynamics. The width of the line is determined by the cosmological time between the beginning of compression and the moment of a sharp increase in the gradient. Within this time interval, the optical thickness of the absorption depends only on the Doppler width and size of the object and is independent on the velocity gradient.

Since the nature of the primary fluctuations at the initial stage is sound waves, the duty cycle of their distribution is close to value $0.5$, and integration along the line of sight, tends to unity, i.e. fluctuations almost densely and evenly cover the picture plane. This mechanism does not depend in any way on the energy release in the Lyman lines or in the X-ray emission from primary stars, nor on the presence of any matter colder than hydrogen.

The era of outbursts of the first stars is known to come after the era of the beginning of compression of primary fluctuations. Therefore, the mechanisms of absorption formation considered in the literature should work at substantially smaller redshifts.
The purpose of this note is to initiate a detailed and comprehensive investigation of the nonlinear stage of compression of primary fluctuations, the study their spectrum and the role of the dark matter.

\end{document}